\documentclass[fleqn,usenatbib]{mnras}
\usepackage{newtxtext,newtxmath}
\usepackage[T1]{fontenc}

\DeclareRobustCommand{\VAN}[3]{#2}
\let\VANthebibliography\thebibliography
\def\thebibliography{\DeclareRobustCommand{\VAN}[3]{##3}\VANthebibliography}
\usepackage{graphicx}	
\usepackage{amsmath}	
\title[Photometric metallicities of RRab stars]{Photometric metallicities of fundamental-mode RR Lyr stars from Gaia $G$-band photometry of globular-cluster variables}

\author[J. Jurcsik, G. Hajdu]{
Johanna Jurcsik,$^{1}$\thanks{E-mail: jurcsik.johanna@gmail.com}
Gergely Hajdu,$^{2}$
\\
$^{1}$ ELKT, CSFKI, Konkoly Observatory, H-1121 Budapest, Konkoly Thege Mikl\'os \'ut 15-17., Hungary\\
$^{2}$Nicolaus Copernicus Astronomical Center, Polish Academy of Sciences, Bartycka 18, 00-716, Warsaw, Poland\\
}

\date{Accepted 2023 August 16. Received 2023 July 06; in original form 2023 May 17.}

\pubyear{2023}

\begin{document}
\label{firstpage}
\pagerange{\pageref{firstpage}--\pageref{lastpage}}
\maketitle

\begin{abstract}
Photometric metallicity formulae of fundamental-mode RR Lyr (RRab) stars are presented using  globular-cluster data exclusively. The aim is to check whether this selection may help  increasing the overall accuracy of the fits and eliminating the systematic bias of the photometric results, namely that they tend to overestimate [Fe/H] of the most metal-poor variables. The $G$-band time-series data available in the Gaia DR3 archive and a new compilation of the published spectroscopic globular cluster [Fe/H] values  on a uniform solar reference metallicity scale are utilized. We have derived a new ${\mathrm{[Fe/H]}}_{\mathrm{phot}}- P,\varphi_{31}$ formula, and have diagnosed that no significant increase in the accuracy of the fit can be achieved using non-linear or multi-parameter formulae. The best result is obtained when different formulae are applied for variables with Oosterhoff-type I and II properties. However, even this solution cannot eliminate  the systematic bias of the results completely. This separation of the variables has also led to the conclusion that the photometric estimates of the [Fe/H] are less reliable for the  Oo-type II variables than for the Oo-type I sample.
Published ${\mathrm{[Fe/H]}}_{\mathrm{phot}}$ values and the results of the available photometric formulae in the Gaia $G$-band  are  compared with the present results. It is found that each of the solutions yields very similar results, with similar accuracy and systematic biases. Major differences are detected only in the zero-points of the [Fe/H] scales, and these offsets are larger than differences in the accepted solar reference values would explain.

\end{abstract}


\section{Introduction}

RRab stars, the fundamental-mode radial pulsators on the horizontal branch in the evolutionary stage of core helium burning, are of great astrophysical importance because their physical properties can be easily estimated from the properties of their light variation. The Bailey diagram,  which traces the variation of the luminosity amplitude of RR Lyrae stars as a function of the  pulsation period, is a fundamental diagnostic for studying RRL stars \citep[as a recent interpretation of the Bailey diagram read][]{bono20}. Several relations between the physical properties and the light-curve parameters of RRL stars has been established based on theoretical and empirical studies. These relations are widely employed in the study of individual variables, as well as their host stellar populations \cite[see e.g.][]{bell20}. 

 The luminosity and the radius of RRL stars correlate with the pulsation period, which can be measured with the highest accuracy out of all physical parameters, and the metallicity \citep[see e.g.][]{mul23,marc15}. The knowledge of the metallicity of stars is crucial for evolutionary and cosmological aspects, too. Therefore, this is not surprising that several  metallicity formulae in different photometric bands has been published after the invention of a very simple, linear [Fe/H]$ - P, \varphi_{31}$ relation by \cite{jk96}.  The accuracy of these photometric metallicities are estimated to be about $0.13-0.16$ dex, typically. The photometric [Fe/H] formulae are widely used to determine globular cluster (GC) metallicities \cite[see e.g.,][]{ferro}, although these formulae systematically overestimate the metallicity of the most metal-poor clusters by $\sim0.2$ dex.
 
 The accuracy of an empirical relation between different observables depends on the accuracy of the input data  that the calibration of the formula is based on, and also on the limits of the inherent physical connection between the compared quantities.  The estimate of systematic and random errors of the observations can be considered reliable in most of the cases, however, the inherent limits of empirical relations is usually unknown.

Concerning the calibration of the photometric metallicity, the most problematic issue is the accepted [Fe/H] values of stars in the calibrating sample. 
The scatter of the different spectroscopic [Fe/H] data of the calibrating stars that are used to establish a photometric metallicity formulae is about 0.15 dex, typically. Therefore, this may account for the similar final accuracy of the predictions. However, what the best possible attainable accuracy is, is an open question for any photometric [Fe/H] formula. 

Regarding field stars, several high-dispersion spectroscopic studies of individual stars and large spectroscopic survey data have been published, but because of the problems with the dynamic atmosphere of RRL stars, the atmospheric parameter determinations, the line list adopted and several other factors, the iron abundance estimates are very divergent in some cases. 

On the other hand, the GC metallicities seem  to have reached a consensus. These [Fe/H] values are determined using high-resolution spectra of giant stars for most of the clusters, and have smaller dispersion for the individual clusters than the [Fe/H] observations of individual field RR Lyrae stars, typically.

Therefore, in this paper, we make an attempt to establish new photometric metallicity formulae using   GC data exclusively, to see whether this would help to improve the overall fitting accuracy  and to reduce the systematic bias  towards the  metal-poor end of the [Fe/H] scale of the predictions. For this purpose we utilize the time-series photometry from Gaia DR3, which, as of today, yields the most homogeneous, complete sample of good-quality  light curves (LCs) of globular cluster variables.

\section{The data}
\subsection{Target selection}
According to our present-day knowledge, most of the GCs host multiple generations of stars of different chemical compositions. 
Several studies of the Hubble Space Telescope UV Legacy Survey of Galactic Globular Clusters \citep{mil15}, which utilizes the pseudo-two-colour diagrams (chromosome maps), indicate that the elemental distributions of the different stellar populations in GCs are divergent. High-resolution spectroscopic studies of individual clusters are in line with this picture.

Spread in the iron content has been also revealed in a few clusters, but its range typically  does not exceed $0.1-0.2$ dex and/or the population of the extreme metallicity stars is marginal in most cases  \citep[see e.g. Table 8 in][]{marino21}.
Moreover, some of the results concerning cluster metallicities are controversial, e.g.,  \cite{carM54} concluded that NGC6715 (M54) in the Sagittarius dwarf galaxy is similar to $\omega$ Cen in its [Fe/H] range and dispersion, but according to the recent analyses of the APOGEE spectra its [Fe/H] spread is found to be smaller  \citep{mesz20,fer21}. \cite{muc16b} have  shown that even the chemical composition analyses of high-resolution spectroscopic (HRS) observations may lead to spurious results concerning the metallicity spread in some cases.

Therefore, we decided to utilize every cluster with at least a single RRab star that have Gaia $G$-band LC in the process of deriving connection between the cluster metallicity and the light-curve parameters of RRab stars.

The only exceptions are: a) the most complex galactic GC, $\omega$ Cen; b) the three Oosterhoff-type III (OoIII) GCs (NGC6304, NGC6388 and NGC6441); and c)
NGC104 (47\,Tuc). The long-period and metal-rich variables in the OoIII clusters and in 47\,Tuc do not yield reliable photometric metallicity values.
 The results on the photometric metallicities of these clusters are discussed in Sect.~\ref{OoIII}.

The catalog of the intrinsic iron-abundance spread of GCs \citep{bai19} lists three clusters with iron dispersion significantly larger than 0.1 dex, namely NGC6273,  NGC6715 and $\omega$ Cen. However, the photometric metallicity of the  RRab stars in NGC6273 and NGC6715 are in good agreement with the mean cluster metallicity values, without showing larger dispersion than in other clusters, therefore these stars are kept in the calibrating sample of the metallicity formula.

\subsection{Spectroscopic data}\label{sect:sp}
Since the 2010 electronic edition of the catalog of GC data \citep[][hereafter H10]{h10,h96} several spectroscopic surveys and detailed chemical composition analyses of stars in globular clusters have been published. The latest compilation of GC metallicities by B. Dias \citep{dias15,dias16a,dias16b,vas18} does not include all the published HRS data of individual GCs. Therefore, to obtain the possible best result, we have collected all  the HRS [Fe/H] measurements, which were not included in the H10 catalog, of the GCs of our interest. As HRS observations are not available for some GCs, the CaII triplet metallicity values  published by \cite{sav12,dias16b,vas18} and \cite{gei23} are also taken into account. 

The [Fe/H] values published by different authors refer to different solar compositions. E.g., the Gaia-ESO Survey \citep{pan17} accepts  log$\epsilon{\mathrm{Fe}_\odot}=7.45$ \citep{solar} solar value, while the H10 [Fe/H] values are matched to the  UVES scale  \citep{C09} adopting  log$\epsilon{\mathrm{Fe}_\odot}=7.54$  \citep{gra03}.

Most of the published [Fe/H] values of RRL stars are matched to the log$\epsilon{\mathrm{Fe}_\odot}=7.50$ \citep{solar2009} value, therefore, in order to obtain the possible most homogeneous globular cluster matallicities all the data used are transposed to this value. This decision results in a +0.04 dex offset of the H10 data. The results published by \cite{mesz20} and \cite{mass19} are 0.064 dex less metal poor than the cluster [Fe/H] values of \cite{C09} and of \cite{pan17}, even when using the same solar reference value \citep[see figure 4. in][]{mesz20}, therefore these data are also corrected for this systematic difference. The  CaII-triplet results
are also matched to the accepted scale  applying 0.02 dex and 0.04 dex corrections on the \cite{dias16b,vas18,gei23} and on the \cite{sav12} data, respectively.

There are several other reasons of systematic differences between the spectroscopic [Fe/H] results. Differences in the model atmospheres, the ionization state of the lines used, the treatment of non-LTE effects, etc., each may lead to systematically different results.  However, to treat all these problems completely is beyond the scope of this paper.   

The [Fe/H] values published by \cite{roed11}, \cite{roed15} and \cite{sobeck11} for NGC6341, NGC4833 and NGC7078, respectively, are about $0.3-0.4$ dex more metal poor than any other published [Fe/H] values of these clusters. Therefore, we omit these values when calculating the cluster mean [Fe/H] values. These are the only available HRS data which are not involved in the averaging.

The [Fe/H] values accepted are the weighted mean values of the H10 data, the additional high-resolution spectroscopic abundance determinations and the low-resolution CaII-triplet equivalent-widths measurement results  using weights of 3, 2 and 1, respectively. Although these weights are arbitrary, taking into account that the H10 catalog incorporates all the measurements prior 2010, its weight of 3 is reasonable.

\begin{figure}
	\includegraphics[width=\columnwidth]{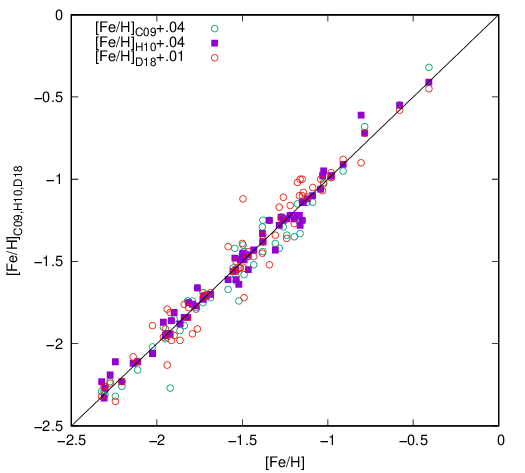}
    \caption{Comparison of the compiled, solar composition standardized (log$\epsilon$(Fe)$_{\odot}=7.50$) [Fe/H] values of GCs with the  metallicities of \citet{C09}, the 2010 edition of the \citet{h96} catalog and the 2nd data release of the compilation of B. Dias \citep{dias15,dias16b,dias16a,vas18}. The latter has an average offset of 0.03 dex compared to the H10 scale.  }
    \label{fig:scale}
\end{figure}

The accepted average [Fe/H] values of the clusters with at least one Gaia $G$-band  RRL LC are listed in Table~\ref{tab:gc_fe}. The H10 metallicities on the accepted solar-composition scale are also given in the table for comparison. The derived mean cluster [Fe/H] values are compared with the \cite{C09} UVES values, the H10 catalog data and the compilation of B. Dias \citep[][]{dias15,dias16a,dias16b,vas18}, each shifted to the same scale accepting log$\epsilon{\mathrm{Fe}}_\odot=7.50$ solar value, in Fig.~\ref{fig:scale}.

\begin{table*}
	\centering
	\caption{Accepted spectroscopic [Fe/H] values of GCs.}
	\label{tab:gc_fe}
	\begin{tabular}{l@{\hspace{1mm}}c@{\hspace{1mm}}c@{\hspace{1mm}}c@{\hspace{1mm}}ll@{\hspace{1mm}}c@{\hspace{1mm}}c@{\hspace{1mm}}c@{\hspace{1mm}}l} 
		\hline
  Cluster  & [Fe/H]$_{\mathrm{mean}}$ & rms& [Fe/H]$^{*}_{\mathrm{H10}}$ &Ref. & Cluster  & [Fe/H]$_{\mathrm{mean}}$ & rms&  [Fe/H]$^{*}_{\mathrm{H10}}$ &Ref. \\
  \hline
  Arp2 & $-$1.710 & $-$ & $-$1.71 & 1                       & NGC6316 & $-$0.408 & 0.046 & $-$0.41 & 1,3,4 \\
Djo2 & $-$0.805 & 0.212 & $-$0.61 & 1,2,3,4,5             & NGC6333 & $-$1.730 & $-$ & $-$1.73 & 1 \\  
IC4499 & $-$1.490 & $-$ & $-$1.49 & 1                     & NGC6341 & $-$2.304 & 0.038 & $-$2.27 & 1,9,10,70 \\  
NGC362 & $-$1.166 & 0.070 & $-$1.22 & 1,6,7,8,9,10,11,12  & NGC6355 & $-$1.383 & 0.053 & $-$1.33 & 1,3,4,71  \\ 
NGC1261 & $-$1.272 & 0.036 & $-$1.23 & 1,13,14,15         & NGC6362 & $-$1.024 & 0.065 & $-$0.95 & 1,72,73   \\
NGC1851 & $-$1.148 & 0.081 & $-$1.14 & 1,7,8,9,10,16,17,18,19 & NGC6366 & $-$0.580 & 0.052 & $-$0.55 & 1,3,4,74,75  \\
NGC1904 & $-$1.553 & 0.023 & $-$1.56 & 1,7,8,9,10         & NGC6401 & $-$1.030 & 0.073 & $-$0.98 & 1,3,4,5   \\
NGC2298 & $-$1.865 & 0.043 & $-$1.88 & 1,21,22            & NGC6402 & $-$1.196 & 0.054 & $-$1.24 & 1,76   \\
NGC2419 & $-$2.113 & 0.019 & $-$2.11 & 1,23,24            & NGC6426 & $-$2.243 & 0.116 & $-$2.11 & 1,3,4,77  \\
NGC2808 & $-$1.088 & 0.062 & $-$1.10 & 1,3,7,8,9,10,25,26,27,28,29,30, & NGC6453 & $-$1.490 & 0.038 & $-$1.46 &1,3,4   \\ 
NGC3201 & $-$1.464 & 0.063 & $-$1.55 & 1,3,9,10,32,33,34,35,36 & NGC6535 & $-$1.814 & 0.078 & $-$1.75 & 1,78  \\
NGC4147 & $-$1.792 & 0.039 & $-$1.76 & 1,37               & NGC6541 & $-$1.770 & $-$ & $-$1.77 & 1   \\
NGC4590 & $-$2.274 & 0.084 & $-$2.19 & 1,3,9,10,38        & NGC6558 & $-$1.162 & 0.106 & $-$1.28 & 1,3,25,79,80  \\
NGC4833 & $-$1.899 & 0.073 & $-$1.81 & 1,7,8,39           & NGC6569 & $-$0.784 & 0.094 & $-$0.72 & 1,3,25,81,82  \\ 
NGC5024 & $-$2.026 & 0.035 & $-$2.06 & 1,9,10,40,41,42    & NGC6584 & $-$1.500 & 0.049 & $-$1.46 & 1,83   \\
NGC5053 & $-$2.205 & 0.029 & $-$2.23 & 1,9,10,42,43       & NGC6626 & $-$1.284 & 0.005 & $-$1.28 & 1,84   \\
NGC5272 & $-$1.469 & 0.019 & $-$1.46 & 1,9,10,44          & NGC6638 & $-$0.910 & $-$ & $-$0.91 &1     \\
NGC5466 & $-$1.922 & 0.051 & $-$1.94 & 1,9,10,40          & NGC6642 & $-$1.178 & 0.048 & $-$1.22 & 1,2,5   \\ 
NGC5634 & $-$1.840 & 0.057 & $-$1.84 & 1,3,4,45           & NGC6656 & $-$1.763 & 0.073 & $-$1.66 &1,3,25,54,85,86  \\   
NGC5824 & $-$1.962 & 0.089 & $-$1.87 & 1,3,25,46,47       & NGC6712 & $-$0.980 & $-$ & $-$0.98 & 1   \\
NGC5897 & $-$1.915 & 0.055 & $-$1.86 & 1,3,48             & NGC6715 & $-$1.496 & 0.087 & $-$1.45 & 1,9,25   \\
NGC5904 & $-$1.260 & 0.030 & $-$1.25 & 1,3,9,10,49,50     & NGC6717 & $-$1.220 & $-$ & $-$1.22 & 1   \\
NGC5946 & $-$1.342 & 0.115 & $-$1.25 & 1,3,4              & NGC6723 & $-$1.042 & 0.081 & $-$1.06 & 1,10,87,88,89   \\
NGC5986 & $-$1.538 & 0.015 & $-$1.55 & 1,51               & NGC6779 & $-$1.940 & $-$ & $-$1.94 & 1   \\
NGC6093 & $-$1.726 & 0.020 & $-$1.71 & 1,52               & NGC6864 & $-$1.150 & 0.108 & $-$1.25 & 1,3,4,90  \\
NGC6101 & $-$1.940 & $-$ & $-$1.94 & 1                    & NGC6934 & $-$1.433 & 0.087 & $-$1.43 & 1,91,92   \\
NGC6121 & $-$1.121 & 0.042 & $-$1.12 & 1,3,9,10,29,53,54,55,56,57,58,5 & NGC6981 & $-$1.380 & $-$ & $-$1.38 & 1 \\  
NGC6139 & $-$1.583 & 0.031 & $-$1.61 & 1,25,64            & NGC7006 & $-$1.544 & 0.081 & $-$1.48 & 1,3,25   \\
NGC6171 & $-$0.980 & 0.028 & $-$0.98 & 1,3,9,10           & NGC7078 & $-$2.357 & 0.124 & $-$2.33 & 1,3,8,9,10,93  \\
NGC6205 & $-$1.512 & 0.023 & $-$1.49 & 1,9,10             & NGC7089 & $-$1.537 & 0.051 & $-$1.61 & 1,7,8,9,10,94,95   \\
NGC6229 & $-$1.307 & 0.117 & $-$1.43 & 1,9,10,65          & NGC7099 & $-$2.324 & 0.099 & $-$2.23 & 1,70,83   \\
NGC6235 & $-$1.240 & $-$ & $-$1.24 & 1                    & Pal13   & $-$1.820 & 0.118 & $-$1.84 & 1,96,97   \\
NGC6266 & $-$1.143 & 0.005 & $-$1.14 & 1,66,67            & Pal2    & $-$1.380 & $-$ & $-$1.38 & 1   \\
NGC6273 & $-$1.686 & 0.058 & $-$1.70 & 1,68,69            & Rup106  & $-$1.520 & 0.114 & $-$1.64 & 1,3 25,98,99,100 \\
NGC6293 & $-$1.950 & $-$ & $-$1.95 & 1                    & Ter8    & $-$2.139 & 0.064 & $-$2.12 & 1,3,4,101   \\  
   \hline
\multicolumn{10}{l}{$^{*}$ The \cite{h10} catalog values are transposed to match the  log$\epsilon{\mathrm{Fe}}_\odot=7.50$  solar composition by adding 0.04 dex.}\\
\multicolumn{10}{l}{References:1 \cite{h10}; 2 \cite{gei21}; 3 \cite{dias16b}; 4 \cite{vas18}; 5 \cite{gei23};6 \cite{car13};}\\
\multicolumn{10}{l}{7 \cite{pan17}; 8 \cite{kov19}; 9 \cite{mesz20}; 10 \cite{hor20}; 11 \cite{var22}; 12 \cite{mon23};}\\
\multicolumn{10}{l}{13 \cite{k-h21}; 14 \cite{marino21}; 15 \cite{mun21}; 16 \cite{car11}; 17 \cite{gra12}; 18 \cite{marino14};}\\
\multicolumn{10}{l}{19 \cite{yong15-1851}; 20 \cite{taut22}; 21 \cite{bae22}; 22 \cite{dias16b}; 23 \cite{coh11-2419}; 24 \cite{muc12};}\\
\multicolumn{10}{l}{25 \cite{sav12}; 26 \cite{marino14-2808}; 27 \cite{car15-2808}; 28 \cite{wang16}; 29 \cite{marino17}; 30 \cite{cab19}; }\\
\multicolumn{10}{l}{31 \cite{carlos23}; 32 \cite{mun13}; 33 \cite{sim13}; 34 \cite{muc15}; 35 \cite{mag18}; 36 \cite{marino19};}\\
\multicolumn{10}{l}{37 \cite{vill16}; 38 \cite{sch15}; 39 \cite{car14-4833}; 40 \cite{lamb15}; 41 \cite{bob16}; 42 \cite{chun20};}\\
\multicolumn{10}{l}{43 \cite{sbor15}; 44 \cite{giv16}; 45 \cite{sbor15}; 46 \cite{roed16}; 47 \cite{muc18};}\\
\multicolumn{10}{l}{48 \cite{koch14}; 49 \cite{lai11}; 50 \cite{gra13}; 51 \cite{john17-5986}; 52 \cite{car15}; 53 \cite{muc11};}\\
\multicolumn{10}{l}{54 \cite{marino11}; 55 \cite{vill11}; 56 \cite{vill12}; 57 \cite{mon12}; 58 \cite{doz13};}\\
\multicolumn{10}{l}{59 \cite{mal14};60 \cite{spite16}; 61 \cite{macl16}; 62 \cite{wang17}; 63 \cite{macl18}; 64 \cite{bra15};}\\
\multicolumn{10}{l}{65 \cite{john17-6229}; 66 \cite{yong14-6266}; 67 \cite{lap15}; 68 \cite{john15}; 69 \cite{john17-6273}; 70 \cite{coh11-7099-6341};}\\
\multicolumn{10}{l}{71 \cite{sou23}; 72 \cite{muc16-6362}; 73 \cite{mass17}; 74 \cite{john16}; 75 \cite{puls18}; 76 \cite{john19};}\\
\multicolumn{10}{l}{77 \cite{hanke17}; 78 \cite{bra17}; 79 \cite{barb18}; 80 \cite{ern18} 81 \cite{val11}; 82 \cite{john18};}\\
\multicolumn{10}{l}{83 \cite{omal18}; 84 \cite{vill17}; 85 \cite{roed6656}; 86 \cite{mck22}; 87 \cite{gra15};}\\
\multicolumn{10}{l}{88 \cite{rojas16}; 89 \cite{crest19}; 90 \cite{kach13}; 91 \cite{marino18}; 92 \cite{marino21}; }\\ 
\multicolumn{10}{l}{93 \cite{wor13}; 94 \cite{yong14-7089}; 95 \cite{lar16}; 96 \cite{bra11}; 97 \cite{koch19}; 98 \cite{vill13};}\\
\multicolumn{10}{l}{99 \cite{frel21}; 100 \cite{luc22}; 101 \cite{car14-ter8}.}
 \end{tabular}
\end{table*}

\subsection{Photometric data}\label{sect:phot}

The $G$-band time-series data of RRab stars published in the third data release (DR3) of the {\it{Gaia}} collaboration \cite{gaia16,gaia22} are utilized. 

The positions of  variable stars given in the C. Clement catalog \citep[][hereafter CC01]{cccat}\footnote{http://www.astro.utoronto.ca/~cclement/cat/listngc.html} are cross-correlated with the Gaia DR3 positions to get the epoch-photometry data of RRLs in GCs.  We find time-series data within 1" position match for 2694 stars. The LCs contain $20-100$ data points, typically. Known RRLs and stars of uncertain variable classification are checked visually one by one to select RRab stars of reliable Fourier solution using the MUFRAN \citep{mufran} program.  
Each LC is fitted by appropriate order Fourier sums.  In order to avoid unreliable bumps of the Fourier fits of gappy and/or poor data, a few  (1-3) artificial points are interpolated to stabilize the solution in some cases. Data points with error flag "1" are removed only if they are indeed discrepant.

The LCs are categorized according to their scatter and the reliability of the fit. Good-quality LCs (Sample A) are fitted with at least 5th-order Fourier sums, and they do not show any considerable scatter or sign of the Blazhko effect. Exceptions are V202 in NGC5272, V44 in NGC6864 and V28 in NGC7089; these stars are long-period, small-amplitude variables with sinusoidal-shape LCs, which  are fitted with 3rd-order Fourier sums. The minimum order of the fit for poor-quality LCs (Sample B) is 4. The LCs of these stars are scattered or gappy, many of them are known to show the Blazhko modulation. Variables with unreliable fits and atypical-shape LCs suggesting other types of variability are not used.

We have removed from the sample variables identified as field stars (f,f?) in the CC01 catalog only if their mean magnitudes are discrepant, because this information in the catalog is not always reliable.

Finally, we selected 526 good- (Sample A) and 396 poor-quality (Sample B) RRab LCs belonging to 70 GCs.  The variables of Sample A belong to 64 GCs.

The distribution of the variables in the clusters is very unequal, there are clusters with only 1-2 RRLs, while LCs of 139 variables are utilized in the GC M3.

\section{Photometric metallicity formulae}\label{sect:fe/h}
To derive  relations between the light-curve parameters of  RRab stars and the mean spectroscopic [Fe/H] values of the clusters only the best-quality LCs (Sample A) are used.

The Gaia light-curve parameters indicate a very strong linear correlation between the [Fe/H] and the period and the $\varphi_{31}$ phase parameter, similarly to the $V$-band results published in \cite{jk96}.  We also detect that the results systematically overestimate the [Fe/H]  towards  metal-poor clusters as in our previous study. Neither the overall accuracy of the fit nor the systematics of the differences could be improved using non-linear and/or multi-parameter formulae.

\begin{figure}
\includegraphics[width=\columnwidth]{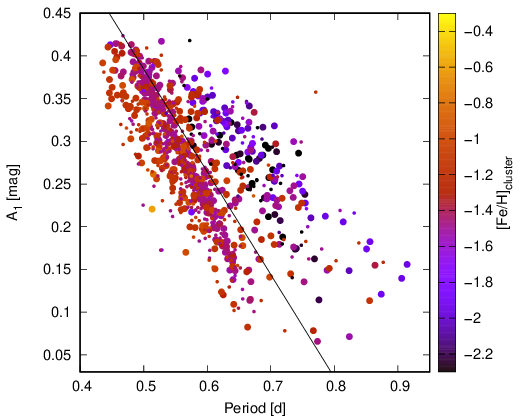}
    \caption{The $Period-amplitude$ (Fourier $A_1$ amplitude) Bailey-diagram of RRab stars of Sample A (large dots) and of Sample B (small dots) is shown. The colour-code denotes the spectroscopic metallicity of the cluster the variable belongs to. The black line in the figure  separates variables supposed to have Oo-type I and Oo- type II properties. }
    \label{fig:oo}
\end{figure}

Different selection criteria are applied to remove the outliers, however, the standard error of the fit cannot be decreased below $0.15-0.16$ dex reliably. It was shown in \cite{j21} that both the $V$-band \citep{jk96} and the $I$-band \citep{smo} metallicity formulae yield systematically different [Fe/H] values for the OoI- and OoII-type variables (which are supposed to be more evolved stars) in the same cluster. 
Therefore, we check the fitting accuracy for Oo-type I and Oo-type II variables separately. Instead of separating the GCs according to  their Oo types, the Oo types of the variables are determined based on their positions on the $period - amplitude$ (Bailey) diagram. 
The  $period - A_1$ plot of the total sample of stars is shown in Fig.~\ref{fig:oo}. Here we use the first-order Fourier amplitude, $A_1$,  as the total amplitude cannot be determined reliably enough for sparse data in some cases.
 It is somewhat arbitrary where to separate the two Oo types, therefore we have tested different linear and quadratic solutions for the separator line. However, no significant differences in the final results emerged depending on the choice of the separator.

We adopt Eq.~\ref{eq:oofit} (shown as black line in Fig.~\ref{fig:oo}) to divide the  Oo types in the total sample of variables used. 
 \begin{equation}
   A_1(G)=  -1.203 P    +  0.986.
	\label{eq:oofit}
\end{equation}

Using   Eq.~\ref{eq:oofit}, Sample A is separated to 318 Oo-type I and 208 Oo-type II variables.

The rms of the residuals of the total Sample A when skipping outliers in successive steps using $3\sigma$ and $2.5\sigma$ thresholds  are 0.160 and 0.149 dex respectively. However, the $3\sigma$ limit is about 0.45 dex, which is too high compared to the full range of the [Fe/H] values of the GCs. Using  $2.5\sigma$ does not improve the fit significantly, but it seems to narrow the range of the residuals artificially, especially when the different Oo types are fitted separately. Therefore, we have decided to use fixed selection limits to cut the outliers. Stars with deviations  larger than 0.6 dex, 0.5 dex and 0.4 dex are removed in successive steps. This process resulted in the selection of $1-3$ percent of the variables as outliers in the different samples.

In the course of checking the outliers to explain their anomalous behaviour, it has been noticed that the photometric metallicity of two of the tree variables in NGC6316 (V15, V18)  underestimate the spectroscopic value significantly, while the Gaia proper-motion values of the third variable (V17) contradict its cluster membership status. Based on HST observations, \cite{deras23} have concluded in a recent paper that the cluster metallicity should be significantly lower than given in the \cite{h10} catalog. No published HRS [Fe/H] data of this cluster is available. Therefore, we decided to remove this cluster from the calibrating sample.

Finally, the following linear  formulae   have been derived. Note, that the $\varphi_{31}$ phase differences correspond to sine series decomposition.

Sample A, 63 GCs, 511 stars, 12 outliers, ${\mathrm{rms}} =0.152$ dex:
\begin{equation}
   {\mathrm{[Fe/H]}}_{\mathrm{phot}}= -5.716(.130) P  +  1.019(.027) \varphi_{31}   -3.504(.097).
\label{eq:fitall}
\end{equation}
Outliers: V54/IC4499,~V11/NGC1851,~V3,V10/NGC5053, 
V3/NGC6171,~V228/NGC6266,~V1/NGC6293, V8/NGC6453, V6/NGC6541, V27/NGC6981, V28/NGC7089, and V4/Pal13.

Sample A, Oo-type I, 30 GCs, 313 stars, 3 outliers, ${\mathrm{rms}} =0.114$ dex:
\begin{equation}
   {\mathrm{[Fe/H]}}_{\mathrm{phot}}= -4.373(.196) P  +  0.736(.031) \varphi_{31}   -2.768(.100).
	\label{eq:fito1}
\end{equation}
Outliers: V3/NGC5053, V6/NGC6541, and V4/Pal13.

Sample A, Oo-type II, 51 GCs, 200 stars, 7 outliers, ${\mathrm{rms}} =0.169$ dex: 
\begin{equation}
   {\mathrm{[Fe/H]}}_{\mathrm{phot}}= -7.212(.248) P  +  1.353(.048) \varphi_{31}   -4.324(.157).
	\label{eq:fito2}
\end{equation}
Outliers: V54/IC4499, V10/NGC5053, V202/NGC5272,  V1/NGC6293,  V8/NGC6453, V27/NGC6981, and V28/NGC7089.

The goodness of the fits using more parameters and/or non-linear formulae are very similar. There is no formula which would yield a  residual scatter better than 0.148 dex  using the data-set Eq.~\ref{eq:fitall} has been defined on.

 The outlier stars are investigated separately to explain their anomalous [Fe/H]$_{\mathrm{phot}}$ values. It is found that: 
\newline
V54/IC4499,~V11/NGC1851,~and~V27/NGC6981 are long-period variables with large amplitudes;
\newline
V10/NGC5053,~V202/NGC5272,~V228/NGC6266,~V28/NGC7089 are also long-period variables but with very low amplitudes;
\newline V3/NGC5053 is the shortest period RRab in the cluster, its total amplitude is 0.2 mag smaller than the amplitude of V1,  the next shortest period RRab at a 0.05~d longer period in this cluster;
\newline V8/NGC6453 is 0.3 mag brighter and its LC shape is markedly different from the similar-period V4 in the cluster;
\newline V3/NGC6171: its amplitude is 0.2 mag smaller than the other Oo-type I variables with similar periods in the cluster;
\newline V4/Pal13: there are two good LCs in the cluster (V3 and V4).  Based on their periods and amplitudes they are  Oo-type I RRLs, however, despite their similar periods, V4 is 0.1 mag fainter than V3 and and its total amplitude is 0.2 mag smaller;
\newline V6/NGC6541: its 0.45 d period is anomalously short for a cluster with [Fe/H]=$-1.77$.
\newline V1/NGC6293: there is no other RRL LC in this cluster to compare.

Summarizing, most of these stars have extremely short or long periods compared to the bulk of the cluster variables, and their amplitudes are anomalously large or small. The others show differing light-curve characteristics and/or mean brightness in comparison with similar period stars in the given cluster. Eight of the 13 selected outliers are Oo-type II variables based on their periods and amplitudes.

\begin{figure}
    \includegraphics[width=9cm]{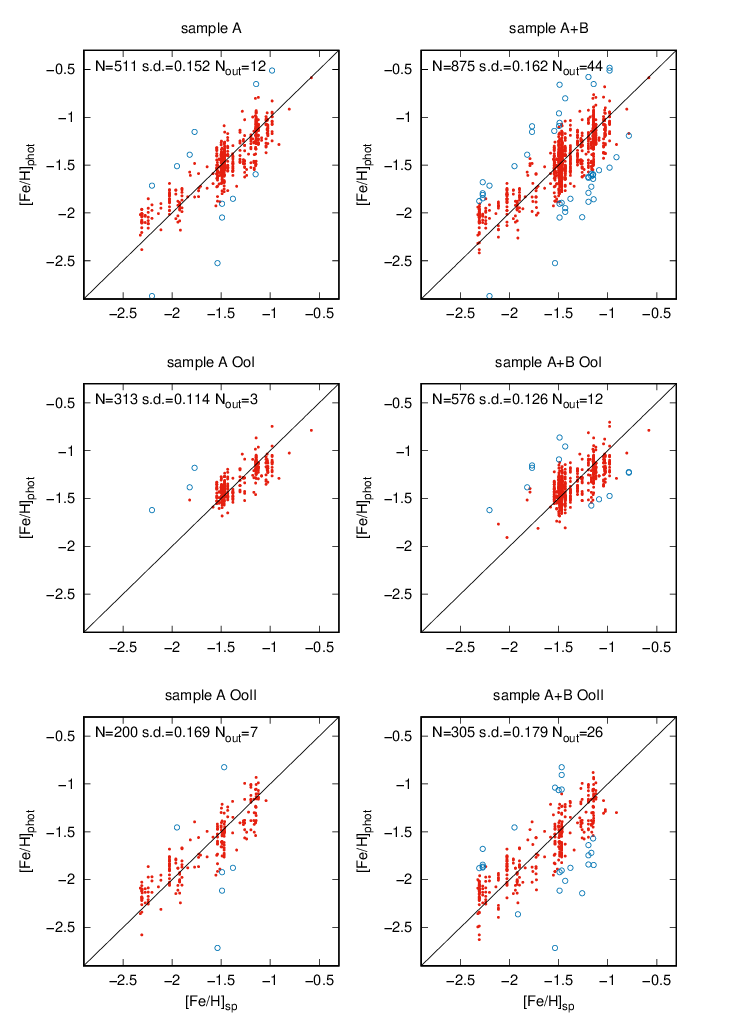}
    \caption{Comparison of the Gaia $G$-band photometric and the spectroscopic [Fe/H] values of globular-cluster RRab stars. The left-hand panels show the results for the best quality LCs (Sample A).  The top-, middle- and bottom-left-hand panels show the results for the complete Sample A, its Oo-type I and Oo-type II sub-samples applying Eqs.~\ref{eq:fitall}, \ref{eq:fito1} and Eq.~\ref{eq:fito2}, respectively.  The results of the combined sample of good and poor quality LCs (Sample A + Sample B) are shown in the right-hand panels of the figure. Removed outliers are indicated by open circles in each plot. The statistical properties of the fits are given near the top of each panel.}
    \label{fig:sig4}
\end{figure}

\begin{figure}
    \includegraphics[width=9cm]{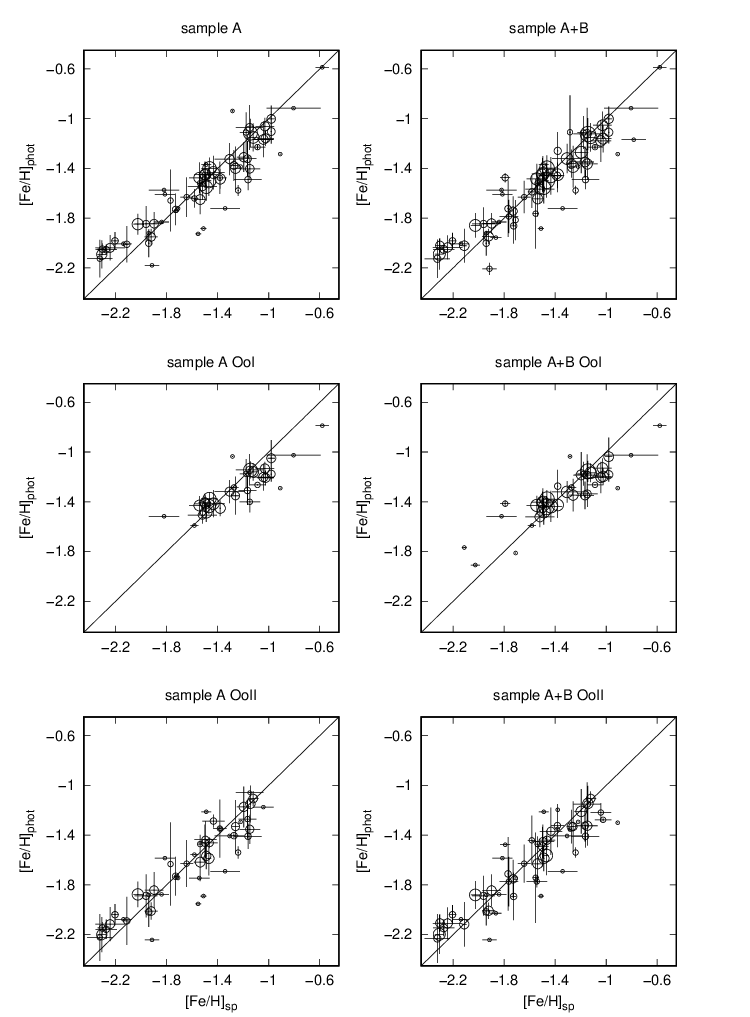}
    \caption{The same panels  are shown for the cluster mean values as in Fig.~\ref{fig:sig4}. The circles size is scaled according to the number of LCs used to derive the mean photometric [Fe/H] value. Error-bars correspond to the dispersion of the ${\mathrm{[Fe/H]}}_{\mathrm{phot}}$ and ${\mathrm{[Fe/H]}}_{\mathrm{sp}}$ values that the means are calculated from.}
    \label{fig:clmean}
\end{figure} 

The photometric metallicities versus the mean cluster spectroscopic [Fe/H] values are shown for Sample A and for Sample A+B in the left-hand and right-hand panels of Fig.~\ref{fig:sig4}. Outliers are shown by open circles, these stars are used neither in deriving the metallicty formulae, nor in calculating the mean cluster values.
The top, middle and bottom panels document the results applying Eqs.~\ref{eq:fitall}, \ref{eq:fito1} and Eq.~\ref{eq:fito2} using the LC parameters of the full samples and on its Oo-type I and Oo-type II sub-samples, respectively. The LCs of Sample B might be biased by Blazhko modulation and/or data sparseness or scatter, this is the reason for the large number of the outliers in these plots. However, the overall agreement with the spectroscopic data is also evident in the right-hand panels of Fig~\ref{fig:sig4}, indicating that the photometric metallicity estimates work relatively well on LCs of poor quality, too.

The Oo-type I sample covers only about the half of the total [Fe/H] range of the clusters, meaning that Eq.~\ref{eq:fito1} is valid only in the $-0.9 > \mathrm{[Fe/H]} > -1.5$ range. However, the significant decrease of the rms of the residuals between Sample A and its Oo-type I part arises not only from the narrow [Fe/H] range but also from the omission of the RRab stars with Oo-type II properties from this sample. The scatter of the OoI-type stars in the middle-left-hand panel of Fig.~\ref{fig:sig4} is smaller than in this part of the plot shown in the top panel.

This is in contrast with the results for the Oo-type II sample (bottom panels in Fig.~\ref{fig:sig4}). The separation of the Oo types does not improve the fitting accuracy in this case, the rms of the fit is even larger than for the total sample. 
This is in line with the outliers preferably being Oo-type II variables. Although Oo-type II variables are less numerous than Oo-type I stars, there are about 2-3 times as many outliers detected in this sample, compared to the Oo-type I variables.  

The Oo-type II sample covers nearly the full metallicity range of the used GCs, and this solution yields a better agreement with the spectroscopic data at the metal-poor end than the results using  Eq.~\ref{eq:fitall}. 
 The mean differences between the  ${\mathrm{[Fe/H]}}_{\mathrm{phot}}$  and ${\mathrm{[Fe/H]}}_{\mathrm{sp}}$ values for the 10 clusters with [Fe/H]$ < -1.95$ dex are 0.19 dex and 0.11 dex when using  Eq.~\ref{eq:fitall} and Eq.~\ref{eq:fito2}, respectively.

The panels in Fig.~\ref{fig:clmean} show the results for the mean cluster values in a similar plot as in Fig.~\ref{fig:sig4}.  The circles size is scaled to reflect the number of variables used to derive the mean value of ${\mathrm{[Fe/H]}}_{\mathrm{phot}}$. The error-bars indicate the rms scatter of the spectroscopic and photometric [Fe/H] values. Note that small (zero) errors do not necessarily indicate high accuracy, because single photometric or spectroscopic data is shown by zero error.

The [Fe/H]$_{\mathrm{phot}}$ values derived for sample A and for the total sample of variables (Sample A + Sample B)  are given in Table~\ref{tab:cl1}.
The metallicities are calculated both  
by using Eq.~\ref{eq:fitall} for each variable of the samples, and by using  Eqs.~\ref{eq:fito1}, and ~\ref{eq:fito2} depending on the Oo-type of the stars.

\begin{figure}
    \includegraphics[width=7.5cm]{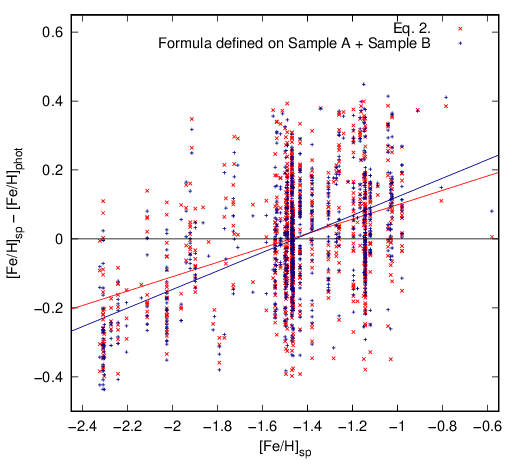}
    \caption{Comparison of the result using Eq.~\ref{eq:fitall},  and according to a formula that is calibrated on the combined sample A+B. The systematic differences of the results are evident in both cases, but its amplitude is smaller when Eq.~\ref{eq:fitall} is applied.}
    \label{fig:1-12comp}
\end{figure}

The mean metallicities derived for the calibrating sample (Sample A) and for the total sample including  also poor-quality LCs (Sample A + Sample B) do not show any significant differences. 
This is partly in line with the results of  \cite{dekany21} and \cite{dekany22}, who concluded  that  the exclusion of the Blazhko stars did not result any improvement of the  photometric metallicity models in the Johnson-Cousins $I$ and the Gaia $G$ photometric bands. However, the benefit of our choice of excluding poor and scattered LCs from the calibrating sample of the metallicity formula is documented in Fig~~\ref{fig:1-12comp}. This figure compares the  ${\mathrm{[Fe/H]}}_{\mathrm{sp}} -  {\mathrm{[Fe/H]}}_{\mathrm{phot}}$ values using Eq.~\ref{eq:fitall} and a similar two parameter formula defined on the combined total sample (Sample A +Sample B). The  systematics of the differences are similar in the two cases, however, when using the combined sample for the calibration the systematic differences tend to be larger than when only the good-quality LCs are utilized.

\begin{table*}
	\caption{Results on the mean photometric [Fe/H] values of the GCs.}
	\label{tab:cl1}
    \begin{minipage}{\textwidth}
	\begin{tabular}{l@{\hspace{2mm}}c@{\hspace{6mm}}c@{\hspace{2mm}}c@{\hspace{2mm}}r@{\hspace{4mm}}c@{\hspace{2mm}}c@{\hspace{2mm}}r@{\hspace{6mm}}c@{\hspace{1mm}}c@{\hspace{2mm}}r@{\hspace{4mm}}c@{\hspace{2mm}}c@{\hspace{2mm}}r}
 \hline
Cluster&&\multicolumn{6}{c}{Sample A}&\multicolumn{6}{c}{Sample A + Sample B}\\
\hline
   & [Fe/H]$_{\mathrm{sp}}$ & ${\mathrm{[Fe/H]}^{\mathrm{Eq.~\ref{eq:fitall}}}_{\mathrm{phot}}}$ &disp. &N&${\mathrm{[Fe/H]}^{\mathrm{Eq.~\ref{eq:fito1},\ref{eq:fito2}}}_{\mathrm{phot}}}$ &disp. &N& ${\mathrm{[Fe/H]}^{\mathrm{Eq.~\ref{eq:fitall}}}_{\mathrm{phot}}}$ &disp. &N&${\mathrm{[Fe/H]}^{\mathrm{Eq.~\ref{eq:fito1},\ref{eq:fito2}}}_{\mathrm{phot}}}$ &disp. &N\\ 
\hline
Arp2     &  $-$1.710  &  $-$1.722  &  $-$      &  1   &  $-$1.744  &  $-$      &  1  &  $-$1.815  &  0.161  &  2  &  $-$1.778  &  0.049  &  2  \\
Djo2     &  $-$0.805  &  $-$0.915  &  $-$      &  1   &  $-$1.026  &  $-$      &  1  &  $-$0.915  &  $-$      &  1  &  $-$1.026  &  $-$      &  1  \\  
IC4499   &  $-$1.490  &  $-$1.551  &  0.107  &  31  &  $-$1.508  &  0.093  &  31 &  $-$1.555  &  0.126  &  47 &  $-$1.510  &  0.130  &  48   \\ 
NGC362   &  $-$1.166  &  $-$1.322  &  0.141  &  5   &  $-$1.294  &  0.115  &  5  &  $-$1.349  &  0.124  &  11 &  $-$1.359  &  0.113  &  12  \\  
NGC1261  &  $-$1.272  &  $-$1.376  &  0.124  &  5   &  $-$1.356  &  0.067  &  5  &  $-$1.361  &  0.113  &  9  &  $-$1.297  &  0.133  &  9   \\ 
NGC1851  &  $-$1.148  &  $-$1.403  &  0.083  &  7   &  $-$1.365  &  0.106  &  8  &  $-$1.364  &  0.123  &  13 &  $-$1.307  &  0.160  &  14  \\  
NGC1904  &  $-$1.553  &  $-$1.927  &  $-$      &  1   &  $-$1.952  &  $-$      &  1  &  $-$1.764  &  0.282  &  2  &  $-$1.637  &  0.445  &  2   \\ 
NGC2298  &  $-$1.865  &  $-$       &  $-$      &  0   &  $-$       &  $-$      &  0  &  $-$1.956  &  $-$      &  1  &  $-$2.028  &  $-$      &  1   \\ 
NGC2419  &  $-$2.113  &  $-$2.010  &  0.146  &  3   &  $-$2.090  &  0.193  &  3  &  $-$2.022  &  0.136  &  10 &  $-$2.090  &  0.209  &  10   \\ 
NGC2808  &  $-$1.088  &  $-$1.228  &  0.018  &  2   &  $-$1.265  &  0.002  &  2  &  $-$1.228  &  0.015  &  2  &  $-$1.265  &  0.002  &  2    \\
NGC3201  &  $-$1.464  &  $-$1.399  &  0.086  &  43  &  $-$1.387  &  0.060  &  43 &  $-$1.398  &  0.104  &  70 &  $-$1.383  &  0.084  &  70   \\ 
NGC4147  &  $-$1.792  &  $-$       &  $-$      &  0   &  $-$       &  $-$      &  0  &  $-$1.475  &  0.037  &  3  &  $-$1.436  &  0.040  &  3   \\ 
NGC4590  &  $-$2.274  &  $-$2.054  &  0.041  &  3   &  $-$2.157  &  0.077  &  3  &  $-$2.051  &  0.035  &  5  &  $-$2.140  &  0.068  &  5   \\ 
NGC4833  &  $-$1.899  &  $-$1.844  &  0.094  &  7   &  $-$1.842  &  0.144  &  7  &  $-$1.843  &  0.084  &  9  &  $-$1.846  &  0.125  &  9   \\ 
NGC5024  &  $-$2.026  &  $-$1.848  &  0.083  &  19  &  $-$1.876  &  0.103  &  19 &  $-$1.859  &  0.099  &  23 &  $-$1.885  &  0.118  &  23  \\  
NGC5053  &  $-$2.205  &  $-$1.982  &  0.071  &  3   &  $-$2.039  &  0.084  &  3  &  $-$1.982  &  0.064  &  3  &  $-$2.039  &  0.084  &  3   \\ 
NGC5272  &  $-$1.469  &  $-$1.499  &  0.122  &  55  &  $-$1.471  &  0.106  &  54 &  $-$1.497  &  0.139  &  138 & $-$1.468  &  0.124  &  135   \\ 
NGC5466  &  $-$1.922  &  $-$1.950  &  0.065  &  7   &  $-$2.010  &  0.069  &  7  &  $-$1.947  &  0.061  &  9  &  $-$2.006  &  0.067  &  9   \\ 
NGC5634  &  $-$1.840  &  $-$1.831  &  $-$      &  1   &  $-$1.877  &  $-$      &  1  &  $-$1.831  &  $-$      &  1  &  $-$1.877  &  $-$      &  1   \\ 
NGC5824  &  $-$1.962  &  $-$1.847  &  0.144  &  4   &  $-$1.888  &  0.173  &  4  &  $-$1.847  &  0.133  &  4  &  $-$1.888  &  0.173  &  4   \\ 
NGC5897  &  $-$1.915  &  $-$2.180  &  $-$      &  1   &  $-$2.243  &  $-$      &  1  &  $-$2.207  &  0.048  &  2  &  $-$2.243  &  $-$      &  1   \\ 
NGC5904  &  $-$1.260  &  $-$1.400  &  0.174  &  12  &  $-$1.341  &  0.177  &  12 &  $-$1.387  &  0.168  &  19 &  $-$1.340  &  0.153  &  19   \\ 
NGC5946  &  $-$1.342  &  $-$1.721  &  $-$      &  1   &  $-$1.690  &  $-$      &  1  &  $-$1.721  &  $-$      &  1  &  $-$1.690  &  $-$      &  1    \\
NGC5986  &  $-$1.538  &  $-$1.519  &  0.064  &  2   &  $-$1.472  &  0.075  &  2  &  $-$1.519  &  0.052  &  2  &  $-$1.472  &  0.075  &  2    \\
NGC6093  &  $-$1.726  &  $-$1.741  &  $-$      &  1   &  $-$1.732  &  $-$      &  1  &  $-$1.862  &  0.144  &  3  &  $-$1.947  &  0.193  &  3    \\
NGC6101  &  $-$1.940  &  $-$2.002  &  0.112  &  3   &  $-$2.017  &  0.121  &  3  &  $-$2.002  &  0.100  &  3  &  $-$2.017  &  0.121  &  3    \\
NGC6121  &  $-$1.121  &  $-$1.150  &  0.114  &  22  &  $-$1.143  &  0.071  &  22 &  $-$1.151  &  0.115  &  28 &  $-$1.155  &  0.081  &  28  \\  
NGC6139  &  $-$1.583  &  $-$1.641  &  0.075  &  2   &  $-$1.574  &  0.025  &  2  &  $-$1.588  &  0.130  &  3  &  $-$1.453  &  0.210  &  3    \\
NGC6171  &  $-$0.980  &  $-$1.001  &  0.108  &  7   &  $-$1.050  &  0.145  &  8  &  $-$1.001  &  0.098  &  9  &  $-$1.028  &  0.165  &  11   \\ 
NGC6205  &  $-$1.512  &  $-$1.884  &  $-$      &  1   &  $-$1.890  &  $-$      &  1  &  $-$1.884  &  $-$      &  1  &  $-$1.890  &  $-$      &  1    \\
NGC6229  &  $-$1.307  &  $-$1.325  &  0.128  &  9   &  $-$1.328  &  0.091  &  9  &  $-$1.321  &  0.118  &  22 &  $-$1.325  &  0.086  &  22    \\
NGC6235  &  $-$1.240  &  $-$1.577  &  0.044  &  2   &  $-$1.539  &  0.050  &  2  &  $-$1.577  &  0.036  &  2  &  $-$1.539  &  0.050  &  2    \\
NGC6266  &  $-$1.143  &  $-$1.115  &  0.115  &  53  &  $-$1.144  &  0.085  &  54 &  $-$1.124  &  0.129  &  87 &  $-$1.158  &  0.102  &  88   \\ 
NGC6273  &  $-$1.643  &  $-$1.630  &  0.159  &  2   &  $-$1.629  &  0.188  &  2  &  $-$1.630  &  0.130  &  2  &  $-$1.629  &  0.188  &  2    \\
NGC6333  &  $-$1.730  &  $-$1.729  &  0.126  &  3   &  $-$1.730  &  0.156  &  3  &  $-$1.744  &  0.120  &  6  &  $-$1.763  &  0.165  &  6    \\
NGC6341  &  $-$2.304  &  $-$2.049  &  0.046  &  3   &  $-$2.146  &  0.046  &  3  &  $-$2.019  &  0.058  &  7  &  $-$2.089  &  0.074  &  7    \\
NGC6355  &  $-$1.383  &  $-$1.475  &  0.130  &  2   &  $-$1.347  &  0.233  &  2  &  $-$1.455  &  0.103  &  3  &  $-$1.308  &  0.178  &  3    \\
NGC6362  &  $-$1.024  &  $-$1.158  &  0.062  &  8   &  $-$1.208  &  0.059  &  8  &  $-$1.161  &  0.110  &  16 &  $-$1.201  &  0.077  &  16   \\ 
NGC6366  &  $-$0.580  &  $-$0.586  &  $-$      &  1   &  $-$0.788  &  $-$      &  1  &  $-$0.586  &  $-$      &  1  &  $-$0.788  &  $-$      &  1    \\
NGC6401  &  $-$1.030  &  $-$1.061  &  0.068  &  9   &  $-$1.133  &  0.054  &  9  &  $-$1.052  &  0.109  &  15 &  $-$1.127  &  0.092  &  15    \\
NGC6402  &  $-$1.196  &  $-$1.309  &  0.119  &  9   &  $-$1.173  &  0.164  &  9  &  $-$1.270  &  0.167  &  24 &  $-$1.210  &  0.185  &  26    \\
NGC6426  &  $-$2.243  &  $-$2.039  &  0.101  &  8   &  $-$2.115  &  0.136  &  8  &  $-$2.035  &  0.096  &  9  &  $-$2.105  &  0.131  &  9    \\
NGC6453  &  $-$1.490  &  $-$1.369  &  $-$      &  1   &  $-$1.212  &  $-$      &  1  &  $-$1.369  &  $-$      &  1  &  $-$1.212  &  $-$      &  1    \\
NGC6535  &  $-$1.814  &  $-$1.608  &  $-$      &  1   &  $-$1.585  &  $-$      &  1  &  $-$1.608  &  $-$      &  1  &  $-$1.585  &  $-$      &  1    \\
NGC6541  &  $-$1.770  &  $-$1.658  &  0.248  &  2   &  $-$1.633  &  0.335  &  2  &  $-$1.724  &  0.228  &  3  &  $-$1.764  &  0.328  &  3    \\
NGC6558  &  $-$1.162  &  $-$1.489  &  0.086  &  3   &  $-$1.409  &  0.102  &  3  &  $-$1.489  &  0.077  &  3  &  $-$1.409  &  0.102  &  3    \\
NGC6569  &  $-$0.784  &  $-$       &  $-$      &  0   &  $-$       &  $-$      &  0  &  $-$1.169  &  $-$      &  1  &  $-$       &  $-$      &  0    \\
NGC6584  &  $-$1.500  &  $-$1.451  &  0.107  &  15  &  $-$1.421  &  0.094  &  15  &  $-$1.453  &  0.138  &  22 &  $-$1.435  &  0.132  &  21   \\ 
NGC6626  &  $-$1.284  &  $-$0.938  &  $-$      &  1   &  $-$1.036  &  $-$      &  1   &  $-$1.108  &  0.296  &  2  &  $-$1.195  &  0.226  &  2    \\
NGC6638  &  $-$0.910  &  $-$1.284  &  $-$      &  1   &  $-$1.290  &  $-$      &  1   &  $-$1.284  &  $-$      &  1  &  $-$1.295  &  0.007  &  2    \\
NGC6642  &  $-$1.178  &  $-$1.111  &  0.162  &  2   &  $-$1.175  &  0.118  &  2   &  $-$1.111  &  0.132  &  2  &  $-$1.175  &  0.118  &  2   \\ 
NGC6656  &  $-$1.763  &  $-$       &  $-$      &  0   &  $-$       &  $-$      &  0   &  $-$1.786  &  0.134  &  2  &  $-$1.774  &  0.200  &  2   \\ 
NGC6712  &  $-$0.980  &  $-$1.104  &  0.097  &  5   &  $-$1.177  &  0.062  &  5   &  $-$1.109  &  0.089  &  6  &  $-$1.185  &  0.058  &  6   \\ 
NGC6715  &  $-$1.496  &  $-$1.447  &  0.100  &  17  &  $-$1.421  &  0.087  &  17  &  $-$1.443  &  0.141  &  30 &  $-$1.407  &  0.131  &  31  \\  
NGC6717  &  $-$1.220  &  $-$1.385  &  $-$      &  1   &  $-$1.294  &  $-$      &  1   &  $-$1.385  &  $-$      &  1  &  $-$1.294  &  $-$      &  1  \\  
NGC6723  &  $-$1.042  &  $-$1.167  &  0.142  &  10  &  $-$1.199  &  0.102  &  10  &  $-$1.179  &  0.168  &  14 &  $-$1.212  &  0.134  &  14   \\ 
NGC6779  &  $-$1.940  &  $-$1.921  &  $-$      &  1   &  $-$1.880  &  $-$      &  1   &  $-$1.921  &  $-$      &  1  &  $-$1.880  &  $-$      &  1   \\ 
NGC6864  &  $-$1.150  &  $-$1.070  &  0.180  &  5   &  $-$1.126  &  0.116  &  5   &  $-$1.121  &  0.208  &  7  &  $-$1.204  &  0.186  &  8  \\  
NGC6934  &  $-$1.433  &  $-$1.433  &  0.134  &  23  &  $-$1.398  &  0.111  &  23  &  $-$1.460  &  0.152  &  43 &  $-$1.443  &  0.145  &  44   \\ 
\end{tabular}
\end{minipage}
\end{table*}
\begin{table*}
\begin{minipage}{\textwidth}
\contcaption{}
	\begin{tabular}{l@{\hspace{2mm}}c@{\hspace{6mm}}c@{\hspace{2mm}}c@{\hspace{2mm}}r@{\hspace{4mm}}c@{\hspace{2mm}}c@{\hspace{2mm}}r@{\hspace{6mm}}c@{\hspace{1mm}}c@{\hspace{2mm}}r@{\hspace{4mm}}c@{\hspace{2mm}}c@{\hspace{2mm}}r}
 \hline
Cluster&&\multicolumn{6}{c}{Sample A}&\multicolumn{6}{c}{Sample A + Sample B}\\
\hline
   & [Fe/H]$_{\mathrm{sp}}$ & ${\mathrm{[Fe/H]}^{\mathrm{Eq.~\ref{eq:fitall}}}_{\mathrm{phot}}}$ &disp. &N&${\mathrm{[Fe/H]}^{\mathrm{Eq.~\ref{eq:fito1},\ref{eq:fito2}}}_{\mathrm{phot}}}$ &disp. &N& ${\mathrm{[Fe/H]}^{\mathrm{Eq.~\ref{eq:fitall}}}_{\mathrm{phot}}}$ &disp. &N&${\mathrm{[Fe/H]}^{\mathrm{Eq.~\ref{eq:fito1},\ref{eq:fito2}}}_{\mathrm{phot}}}$ &disp. &N\\ 
\hline
NGC6981  &  $-$1.380  &  $-$1.484  &  0.079  &  14  &  $-$1.445  &  0.066  &  14  &  $-$1.451  &  0.139  &  22 &  $-$1.408  &  0.119  &  22   \\ 
NGC7006  &  $-$1.544  &  $-$1.476  &  0.124  &  16  &  $-$1.450  &  0.107  &  16  &  $-$1.485  &  0.122  &  27 &  $-$1.454  &  0.121  &  28   \\ 
NGC7078  &  $-$2.309  &  $-$2.092  &  0.113  &  13  &  $-$2.199  &  0.141  &  13  &  $-$2.089  &  0.125  &  24 &  $-$2.196  &  0.170  &  24  \\  
NGC7089  &  $-$1.537  &  $-$1.651  &  0.116  &  11  &  $-$1.617  &  0.153  &  11  &  $-$1.644  &  0.144  &  13 &  $-$1.642  &  0.170  &  12  \\  
NGC7099  &  $-$2.324  &  $-$2.126  &  0.151  &  2   &  $-$2.222  &  0.188  &  2  &  $-$2.127  &  0.153  &  4  &  $-$2.235  &  0.229  &  4   \\ 
Pal13    &  $-$1.820  &  $-$1.574  &  $-$      &  1   &  $-$1.517  &  $-$      &  1  &  $-$1.574  &  $-$      &  1  &  $-$1.517  &  $-$       &  1   \\ 
Pal2     &  $-$1.380  &  $-$       &  $-$      &  0   &  $-$       &  $-$      &  0  &  $-$1.259  &  0.150  &  4  &  $-$1.255  &  0.114  &  4   \\ 
Rup106   &  $-$1.520  &  $-$1.547  &  0.102  &  6   &  $-$1.507  &  0.068  &  6  &  $-$1.567  &  0.117  &  11 &  $-$1.529  &  0.092  &  11   \\ 
Ter8     &  $-$2.139  &  $-$2.006  &  $-$      &  1   &  $-$2.076  &  $-$      &  1  &  $-$2.006  &  $-$      &  1  &  $-$2.076  &  $-$      &  1   \\ 
\hline
\end{tabular}
\end{minipage}
\end{table*}

\section{Results on peculiar clusters}\label{pec}
\subsection{$\omega$ Cen}
\begin{figure}
\begin{center}
    \includegraphics[width=7.5cm]{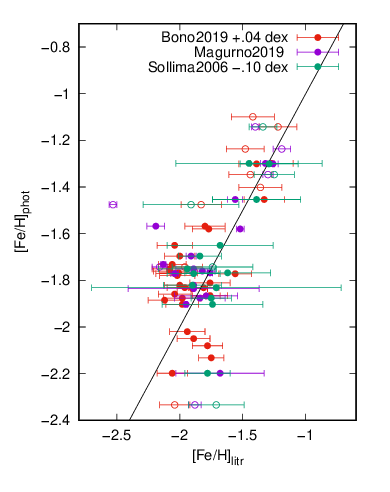}
    \end{center}
    \caption{Comparison of the photometric and the literature [Fe/H] values of RRab stars in $\omega$ Cen. The literature data are taken form the spectroscopic studies of \citet{soll} and \citet{mag19} and are the values derived from $period-luminosity-metallicity$-relation \citep{bono19}. The ${\mathrm{[Fe/H]}}_{\mathrm{phot}}$ values are calculated according to Eq.~\ref{eq:fitall}. Open symbols refer to the results obtained from poor-quality LCs. The black line is the identity function. Taking into account the uncertainties, the agreement is satisfactory. The 0.04 dex offset of the \citet{bono19} data corresponds to the differences of the solar reference values used. There is a  $-0.1$ dex mean difference in comparison with the results  of the data published by \citet{soll} that should have some other origin.    
  }
    \label{fig:ocen}
\end{figure}

We have found  31 and 8 good- and poor-quality LCs in $\omega$ Cen, respectively, in the Gaia DR3  $G$-band time-series data, which are suitable to estimate their ${\mathrm{[Fe/H]}}_{\mathrm{phot}}$ values. The results are compared with the spectroscopic [Fe/H] measured by \citet{soll} and \citet{mag19}, and with the metallicities estimated from $period-luminosity-metallicity$-relation by \cite{bono19} in Fig.~\ref{fig:ocen}.
The 0.00 and 0.04 dex overall shifts between the data correspond to the differences between the reference solar iron abundances for the  \citet{mag19} and \cite{bono19} data, respectively, while the $-0.1$ dex shift of the [Fe/H] values given in \citet{soll} refers to the mean difference obtained from the data. 

Taking into account the large uncertainties of most of the literature data, and the uncertainty of the LC parameters of some of the RRLs we use, the agreement between our and previous results is satisfactory.
Significant differences are detected only in a few cases, and the differences between the literature data of the same star (these data have the same ${\mathrm{[Fe/H]}}_{\mathrm{phot}}$ values, i.e., they lay along a horizontal line in Fig.~\ref{fig:ocen}) are of similar size to their differences from the photometric [Fe/H] values.

\begin{figure}
    \includegraphics[width=\columnwidth]{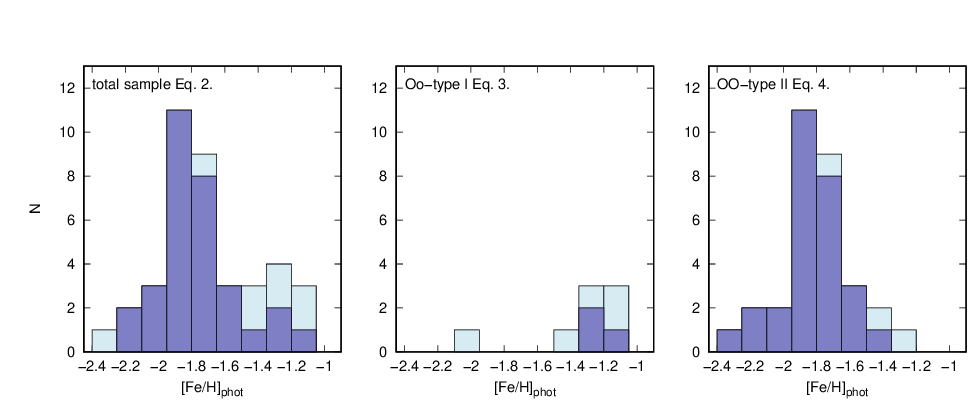}
    \caption{The histograms of the derived ${\mathrm{[Fe/H]}}_{\mathrm{phot}}$ values for $\omega$~Cen  RRab stars are shown. The dark- and the light-blue columns correspond to the results obtained for the good- and the poor-quality LCs, respectively. The left-hand panel documents the distribution of the  ${\mathrm{[Fe/H]}}_{\mathrm{phot}}$ values for the total sample of the variables in the cluster using Eq.~\ref{eq:fitall}. The middle and right-hand panels are the results for the OoI- and OoII-type stars calculated according to Eqs.~\ref{eq:fito1} and \ref{eq:fito2}, respectively.} 
    \label{fig:ocenhist}
\end{figure}

The distributions of the obtained metallicty values using  Eqs.~\ref{eq:fitall},\ref{eq:fito1} and \ref{eq:fito2} on the total, the OoI, and the OoII sub-samples, respectively, are shown in the panels of Fig~\ref{fig:ocenhist}.
The [Fe/H] bins around  $-1.90$, $-1.30$ and $-1.90$ are the most populated bins in the three samples, respectively. 

 The [Fe/H] distribution shown for the total sample in Fig.~\ref{fig:ocenhist} is very similar to the distribution detected from spectroscopic measurements by \citet[][see their figure 8]{mag19}. There is a prominent peak at around $-1.8 - -1.9$ dex and both the more metal-poor and less metal-poor tails are populated. The photometric results also reveal that the higher metallicity RRLs are Oo-type I stars (the LC of the only exception is uncertain), while RRLs in the most metal-poor group are Oo-type II variables. The main sample of the variables belongs to the OoII class.

\subsection{Oo-type III clusters and 47 Tuc}\label{OoIII}

 Photometric studies of RRLs in Oo-type III GCs showed that these 
long-period and metal-rich variables do not yield reliable photometric metallicity values using formulae calibrated on the LCs of galactic field RRLs \citep[see e.g.,][]{pritzl02}. Therefore, it is an open question whether these stars obey another photometric metallicity relation, or there is no connection between their [Fe/H] and LC parameters.

 Gaia DR3 provides  LCs of RLs in NGC6304, NGC6388 and NGC6441, however, many of them are of very poor quality and/or their Gaia proper motion values contradict their cluster membership \citep{gaia22,vas21}. The probable cluster members in NGC6441 with good- and poor-quality Gaia LCs are V37,V38,V39,V96 and V43,V46,V52,V61,V66, respectively. The shapes of the LCs of V96 and V52 are, however atypical for RRL stars. The probable members in NGC6388 are V21,V28 (good) and V22, V58 (poor) but the membership status of V28 and V22 are questionable. We did not find Gaia RRL time-series data of any probable clusters member in NGC6304.

Accordingly, these data are not adequate neither in number nor in quality to establish a relation between the cluster metallicities and the LC parameters. 

Moreover, another problem arises from that, that the ${\mathrm{[Fe/H]}}_{\mathrm{sp}}$ values of these clusters are  very similar $-0.41$, $-0.51$ and $-0.42$ for NGC6304, NGC6388 and NGC6441, respectively \citep[][transformed to log$\epsilon$(Fe)$_{\odot}=7.50$]{h10}. As a result,  a direct approach to find an appropriate formula to fit their [Fe/H] would lead to degeneracy, yielding coefficients of the LC parameters   close to zero and a constant term equalling the mean metallicity of the clusters.

 The only RRL star in 47 Tuc, V9, a long-period and large-amplitude variable also in a metal-rich cluster, which however is about 0.3 dex less metal-rich than the three Oo-type III clusters are,  was proposed to be a close analogue to the RRLs in NGC6388 and NGC6441 by \cite{pritzl00}. The photometric [Fe/H] derived for V9 using any of the formulae defined in Sect.~\ref{sect:fe/h} is $0.8-1.0$ dex more metal-poor than the spectroscopic metallicity of the cluster. 
 Trying to find a common [Fe/H] - LC parameter solution using V9/47Tuc and the OoIII-type RRLs has, however, also failed. 

Supposing that some of the outliers listed in Sect~\ref{sect:fe/h} might also belong to the OoIII type, it is checked whether the addition of these stars would help to fit the ${\mathrm{[Fe/H]}}_{\mathrm{sp}}$  with LC parameters. However, this trial has not led to success either.

Thus, we conclude that in order to answer the question whether there is  any photometric metallicity formula  valid for these peculiar RRL stars, more data on a wider [Fe/H] range of similar variables are needed. Unfortunately, there are very few long-period but relatively metal-rich field RRab stars with high-resolution spectroscopic [Fe/H] determinations (e.g., XZ Gru, HK Pup, and AX Leo), and the similarity of these stars to the variables in the OoIII clusters might be questioned. Moreover, the significant differences between the populations of the Oosterhoff types in the Galactic field and GC samples as discussed by e.g., \cite{fabrizio21} warns that the Oo classification scheme may not be hold for Galactic field stars. Therefore, to extend the OoIII type sample significantly is unrealistic at present.

\section{Comparison with other [Fe/H]$_{\mathrm{phot}}$ results}

\begin{figure}
    \includegraphics[width=9.2cm]{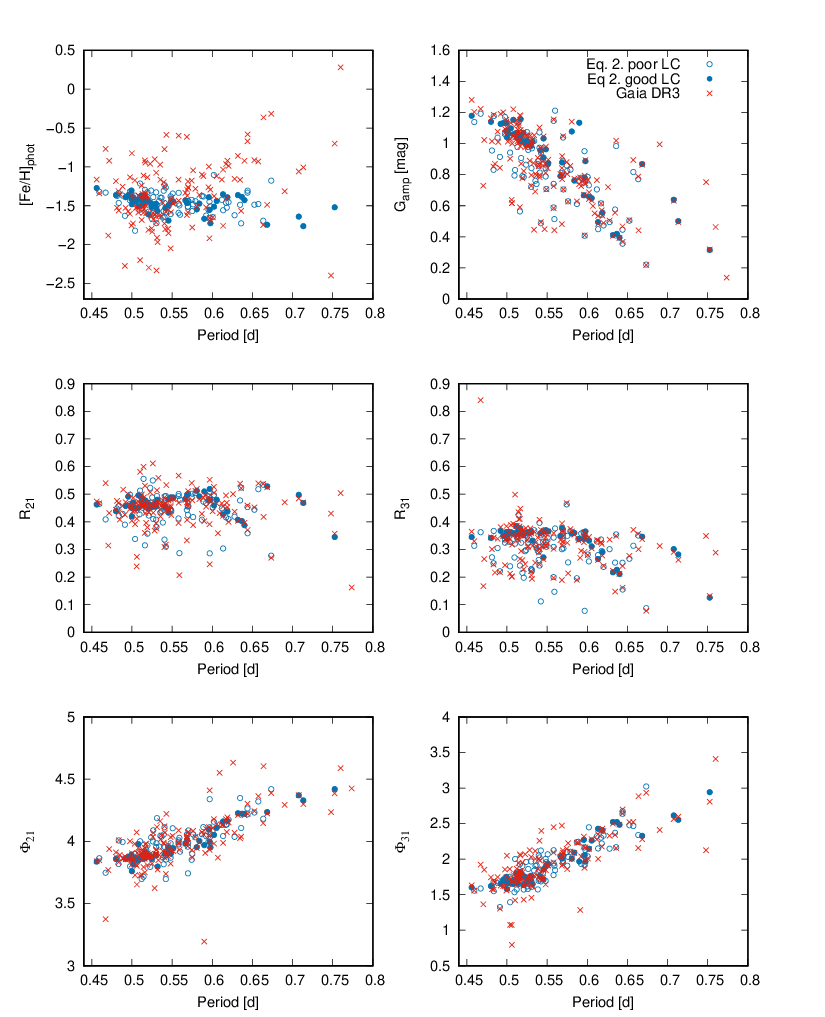}
    \caption{Comparison of the ${\mathrm{[Fe/H]}}_{\mathrm{phot}}$ values and the LC parameters of RRab stars in NGC5272 (M3) derived in this paper (filled and open blue circles for good- and poor-quality LCs, respectively) and as given by the Gaia DR3 RR Lyrae pipeline \citep{gaiarr} (red crosses). } 
    \label{fig:dr3comp}
\end{figure}

\begin{figure*}
    \includegraphics[width=17.8cm]{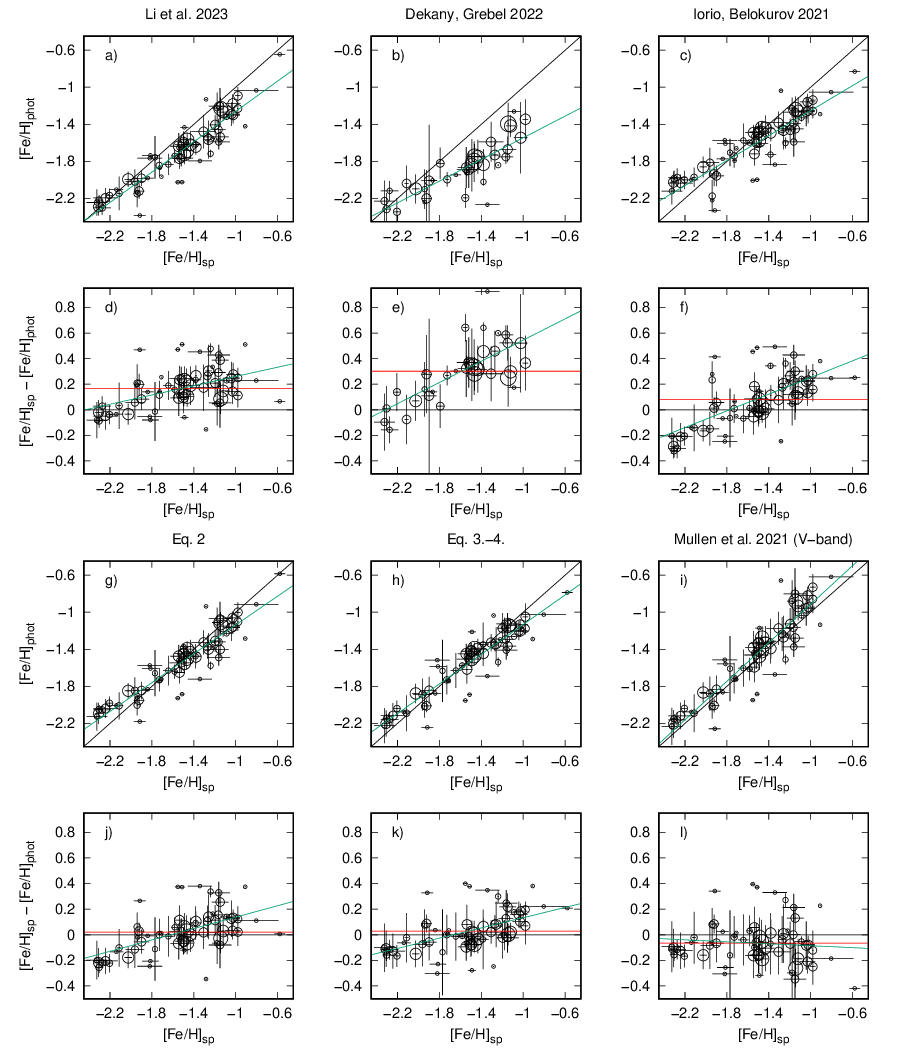}
    \caption{Comparison of the  clusters'  mean   ${\mathrm{[Fe/H]}}_{\mathrm{phot}}$ values derived from the Gaia photometric data using eq. 1. of \citet{li23}, the [Fe/H] values published by \citet{dekany22}, calculated according to eq. 3. of \citet{iorio21}, and Eq.~\ref{eq:fitall}, and Eqs. \ref{eq:fito1},\ref{eq:fito2} of this paper  with their  spectroscopic metallicities (${\mathrm{[Fe/H]}}_{\mathrm{sp}}$) are shown in the panels of $a),b),c),g),h)$, respectively,  while panel $i)$ plots the $V$-band ${\mathrm{[Fe/H]}}_{\mathrm{phot}}$ results using eqs. 6. and 9. of \citet{mull21} after transforming our $G$-band $\varphi_{31}$ values to $V$-band.
    The circles size is scaled according to the number of the LCs used in the cluster, the horizontal and vertical error-bars indicate the scatter of the ${\mathrm{[Fe/H]}}_{\mathrm{sp}}$ and the  ${\mathrm{[Fe/H]}}_{\mathrm{phot}}$ values. The $d),e),f),j),k),l)$ panels document the differences between the  ${\mathrm{[Fe/H]}}_{\mathrm{sp}}$ and the mean ${\mathrm{[Fe/H]}}_{\mathrm{phot}}$ values. The black lines show the identity or the zero functions, while the green and red lines correspond to linear or constant fits to the data. Results obtained using the stars of Sample A are plotted in this figure, with the exception of panes $b)$, and $e)$.   }
    \label{fig:clmeancomp3}
\end{figure*}

The RR Lyr light-curve parameters and metallicities available in the Gaia DR3  archive are compared with the LC parameters and the ${\mathrm{[Fe/H]}}_{\mathrm{phot}}$ values derived in Sect.~\ref{sect:fe/h} for the globular cluster NGC5272 (M3) in Fig.~\ref{fig:dr3comp}. 

The cross-matching of the Gaia DR3 epoch photometry data with the CC01 catalog of variables provides LCs of 163 RRab stars in M3. As a result of the one by one analysis we selected 55 good- and 84 poor-quality LCs from this sample and no reliable Fourier solution was obtained for 24 stars. The Fourier parameters and the derived [Fe/H] values using Eq.~\ref{eq:fitall} of the good and the poor LCs are shown by filled and open blue circles in Fig.~\ref{fig:dr3comp}, respectively. For comparison, the LC parameters and the metallicities of 158 RRab stars provided by the Gaia DR3 \citep{gaiarr} are plotted as cross marks. These ${\mathrm{[Fe/H]}}_{\mathrm{phot}}$ values are calculated according to the non-linear formula of \cite{nemec} transformed to the Gaia $G$-band magnitudes. 

The largest differences between the DR3 and the recent results are in the photometric metallicities. The \cite{nemec} formula yields significantly less reliable metallicities than the formualae given in Sect.~\ref{sect:fe/h}. The differences in the peak to peak amplitude and the Fourier parameters are less significant, part of the differences arises from the omission  of 24 LCs with unreliable Fourier solutions in our sample, but the parameters of many stars in common differs also, especially for noisy or Blazhko-star data (poor LCs). The Gaia pipeline detects most of the RRab stars correctly, but e.g, V42, a non-modulated large-amplitude (1.1 mag) variable with a good quality epoch photometry available in Gaia DR3, is not included in the catalog of RR Lyrae variables  \citep{gaiarr}. The different number of RRab stars in our and in the Gaia DR3 RR Lyrae samples indicates that there are 5 RRab variables that were not identified as such by the RR Lyrae pipeline.

Photometric metallicities of RRL stars using Gaia $G$-band light curves  of the DR2 and DR3 archives were published in \cite{iorio21}, \cite{dekany22}, and \cite{li23}. These results for globular cluster RRab stars and the  ${\mathrm{[Fe/H]}}_{\mathrm{phot}}$ values calculated according to Eq.~\ref{eq:fitall}, and   Eqs.~\ref{eq:fito1}-\ref{eq:fito2} are compared with the spectroscopic cluster metallicities listed in Table \ref{tab:gc_fe} in Fig.~\ref{fig:clmeancomp3}.  For completeness, we also show the results obtained using one of the recent $V$-band formulae \citep{mull21}. However, we note that the comparison of  from different bands  needs additional transformations of the involved parameters, which might not be known accurately enough.

Using the  parameters of the 508 good-quality LCs (Sample A minus the outliers listed in Sect.~\ref{sect:fe/h}) the clusters' mean ${\mathrm{[Fe/H]}}_{\mathrm{phot}}$ values are calculated according to the formulae derived by \cite{iorio21}  and \cite{li23}  in panels a) and c). \cite{dekany22} did not give an explicit metallicity formula, instead they published the  ${\mathrm{[Fe/H]}}_{\mathrm{phot}}$ values of over 60,000 Galactic RRL stars. Cross-correlating their catalog with the CC01 catalog  we find 618 common stars. Panel b) in Fig.~\ref{fig:clmeancomp3} displays the mean cluster   ${\mathrm{[Fe/H]}}_{\mathrm{phot}}$ values calculated from the  metallicties published by  \cite{dekany22} for these stars. The results using Eq.~\ref{eq:fitall}  and by applying Eq.~\ref{eq:fito1} and Eq.~\ref{eq:fito2} on the LC parameters of Oo-type I and II variables separately are shown in panels g) and h).  Panels d),e),f) and j),k)  show the differences between the mean
 ${\mathrm{[Fe/H]}}_{\mathrm{phot}}$ and ${\mathrm{[Fe/H]}}_{\mathrm{sp}}$  cluster values of the respective data.
The panels, $i)$ and $l)$, document the results using eq. 6 of \cite{mull21}  
after applying the transformation between the $V$ and $G$-band $\varphi_{31}$ parameters according to the formula given in \cite{clementini16}.
The black lines show the identity or the zero functions, while the green and red lines correspond to linear and constant fits to the data, respectively. 

Notwithstanding that the calibrating samples of both the LC parameters and the [Fe/H] values, which the different Gaia-band photometric metallicity determinations are based on, are different, the estimated mean cluster metallicities show consistent results, with the exception of the overall zero-point differences between the results. 
Each ${\mathrm{[Fe/H]}}_{\mathrm{phot}}$ estimate tends to over-  and to under-estimate the [Fe/H] towards the metal-poor and metal-rich ends, respectively, when the mean zero-offsets are taken into account [see $d),e),f),j),k$) panels  in Fig.~\ref{fig:clmeancomp3}].
This systematic bias of the results is the smallest in panels $d)$ and $k)$ i.e., when the 3 parameter formula of \cite{li23}, and when different formulae for the  different Oo-type variables  (Eqs.~\ref{eq:fito1},~\ref{eq:fito2}) are applied. However, none of the $G$-band methods can eliminate this trend of the residuals completely.  On the other hand, the $V$-band formula \citep{mull21} does not show any systematic bias of the results, but an increased scatter in the [$-1.0:-1.6$] [Fe/H] interval is evident in this case. This latter may be the consequence of that simple linear/constant transformations between the phase-parameters in different photometric bands might possess additional uncertainties. 

Except an overall shift, the results obtained by using eq. 1. of \cite{li23} [panels $a),d)$], and Eq.~\ref{eq:fitall} and Eqs.~\ref{eq:fito1},~\ref{eq:fito2} [panels $g),j),h),k)$] seems to be the most compatible with each other, although the formulae are quite different. \cite{li23}  established a 3 parameter ($p,\varphi_{31},R_{21}$) formula, while we could not improve the fitting accuracy by using any additional parameter.

The calibration of the formula published by \cite{li23}  relies on the [Fe/H] values of the \cite{liu20} catalog and the LC parameters given in the Gaia DR3 catalog. Their calibrating sample consists of about 2000 stars with metallicity distribution similar to the distribution in our GC sample. The [Fe/H] of the overwhelming majority of their stars fall in the $-1$ -- $-2$ metallicity range \citep[see figure 2. in][]{li23}. The \cite{liu20} catalog values are  calibrated to field and globular cluster data adopting  log$\epsilon{\mathrm{Fe}}_\odot=7.50$ and log$\epsilon{\mathrm{Fe}}_\odot=7.54$ solar abundances,  respectively.
As a consequence of our choice of the log$\epsilon{\mathrm{Fe}}_\odot=7.50$ solar reference value there should be a $\sim0.02$ dex offset between the results of the formulae published by \cite{li23} and  derived in Sect~\ref{sect:fe/h}. However, as it can be seen in Fig.~\ref{fig:clmeancomp3} there is about   0.17 dex difference between the results, which cannot be  explained by the difference of the solar reference values. As the metallicity distribution of the two calibrations are similar, this also cannot explain the differences of the means.

The [Fe/H] data obtained by \cite{dekany22} [panels b) and e) in Fig.~\ref{fig:clmeancomp3}] show an even larger offset, the average ${\mathrm{[Fe/H]}}_{\mathrm{sp}}-{\mathrm{[Fe/H]}}_{\mathrm{phot}}$ difference is 0.30 dex. This data is based on the photometric metallicities derived from an $I$-band metallicity formula \citep{dekany21}. The spectroscopic [Fe/H] data behind the \cite{dekany22} results were homogenized to match the metallicities published by \cite{Cres21}, and this work accepted  log$\epsilon{\mathrm{Fe}}_\odot=7.50$ solar value.
Consequently, the difference in the solar reference values does not explain the offset between the results. Besides the large offset, the scatter and the amplitude of the systematic of the differences of the mean ${\mathrm{[Fe/H]}}_{\mathrm{sp}}-{\mathrm{[Fe/H]}}_{\mathrm{phot}}$ cluster metallicities [see panel e)] for their [Fe/H] values is also larger than for the other [Fe/H] estimates. This might be because of the poorer quality of the LCs of the Gaia DR2 than in the DR3 archive and also because that the $I$-band photometric metallicities they used for  calibration are, most probably,  already affected by some systematic bias.

Eq. 3. of \cite{iorio21} involves only the period and $\varphi_{31}$ in the fitting process similarly to the relations presented here and they use the DR2 LCs. Their results, which are calibrated to the [Fe/H] values of \cite{lay94} is the closest in zero-point to our accepted scale, however, as these [Fe/H] values are calibrated to an in-homogeneous sample of spectroscopic results and on the Zinn \& West metallicities of globular clusters \citep{zw84}, it cannot be directly established, how much part of this offset comes from the differences in the reference solar values.
Despite the same parameters involved, and the similarity of the results, their coefficients differ significantly from the coefficients of   Eq.~\ref{eq:fitall}.  The scatter of the results and the range of the residual differences of the  \cite{iorio21} formula is only marginally larger than for the results of \cite{li23} and of the formulae given in Sect.~\ref{sect:fe/h}.

\section{Summary and conclusions}

New ${\mathrm{[Fe/H]}}_{\mathrm{phot}}$ formulae of RRab stars in the Gaia $G$-band are derived utilizing globular cluster data exclusively. The calibration of the formulae relies on a new compilation of the spectroscopic [Fe/H] values of the clusters involved in the study. The clusters' [Fe/H] values are unified, adopting  log$\epsilon{\mathrm{Fe}}_\odot=7.50$ as the solar reference value.

Similarly to many previous studies the period and the $\phi_{31}$ epoch-independent phase difference are the LC parameters that yield the best estimate of the [Fe/H]. The fitting accuracy could not be improved using non-linear or multi-parameter formulae.  Based on the detection that the linear $V$- and $I$-band [Fe/H] formulae yield systematically different results for Oo-type I and II variables \citep{j21}, formulae have also been derived for the two Oo types separately. The fitting accuracy of the results could not be increased  significantly this way, but the systematic $\sim0.2$ dex bias at the low-metallicity end is reduced to $\sim0.1$ dex. This procedure has also revealed that the scatter of the results for Oo-type II stars is larger than for the compete sample, additionally, most of the variables omitted from the fits because of their anomalous ${\mathrm{[Fe/H]}}_{\mathrm{phot}}$ values belong to this class of variables.

It seems that the Oo-type II sample of RRab stars does not follow  any photometric metallicity relation strictly. Although no general census on the the explanation of the Oosterhoff dichotomy 
has been reached yet \citep[see e.g.][]{catelan09}, horizontal-branch evolution and morphology
are some of the main issues that influence the structure of the period-amplitude diagram.
In an early study of synthetic horizontal-branch evolution models, 
\cite{ldz} showed that, in part, {\it evolution away from the ZAHB will play a role 
in the Oosterhoff group II clusters by increasing the mean luminosity and lowering the 
mean mass of the stars in the instability strip over the values predicted by ZAHB models.} 
The examination  of the period-amplitude relation by \cite{cs99} supported the hypothesis 
{\it that most RR Lyrae variables in Oosterhoff type I clusters are ZAHB objects while 
those in the Oosterhoff type II clusters are more evolved.}
\cite{bono20} concluded  that {\it evolutionary effects take account of the 
vertical structure (dispersion in amplitude) of the Bailey diagram based on radial 
velocity amplitudes.} As we separated Oo-type I and II stars based on their location 
on the period-amplitude plane, it is plausible to suppose  that variables with significantly 
different amplitudes at fixed period values are in different evolutionary stages.

Therefore, we surmise that Oo-type II stars, most probably, have already evolved off the zero age horizontal branch (ZAHB), and their physical structures may show larger diversity than  the structures of RRLs close to the ZAHB evolutionary stage. This might be the reason for the decreased efficiency of the photometric metallicity formula for Oo-type II variables than for Oo-type I stars.

The inherent accuracy of any simple photometric [Fe/H] formula may be set by the heterogeneity of the variables. We think that the more homologous the variables in a sample, the more accurate ${\mathrm{[Fe/H]}}_{\mathrm{phot}}$ formula is valid for that sample.

The comparison of our [Fe/H] estimates with published Gaia $G$-band photometric metallicities of RRab stars has also led to the conclusion that none of the photometric metallicity formula  valid in this band yields similarly accurate [Fe/H] values all along the possible metallicity range and for variables evolved off the ZAHB. Involving extra parameters in the fits or the application of a complex deep learning method \citep{dekany22} does not seem to improve the results (see Fig.~\ref{fig:clmeancomp3}) substantially. 

We have detected significant differences between the zero points of the different photometric metallicity calibrations, which cannot be attributed to the differences between the accepted solar reference [Fe/H] values. This problem warns, that accurate reference [Fe/H] values on a homogeneous scale are higly needed. The compilation of the GC [Fe/H] data given in Sect.~\ref{sect:sp} aims to help in this problem in part.

\section*{Acknowledgements}
This work has made use of data from the European Space Agency (ESA) mission
{\it Gaia} (\url{https://www.cosmos.esa.int/gaia}), processed by the {\it Gaia}
Data Processing and Analysis Consortium (DPAC,
\url{https://www.cosmos.esa.int/web/gaia/dpac/consortium}). Funding for the DPAC
has been provided by national institutions, in particular the institutions
participating in the {\it Gaia} Multilateral Agreement. This research has been supported by the Hungarian National Research, Development and Innovation Office (NKFIH) grants NN-129075 and K-129249. The research leading to these results has received funding from the European Research Council (ERC) under the European Union’s Horizon
2020 research and in novation program (grant agreement No. 695099).

\section*{Data Availability}
The paper utilizes data of the Gaia DR3 archive, and published [Fe/H] values of globular clusters, all of which are publicly available.



\bibliographystyle{mnras}
\bibliography{met} 








\bsp	
\label{lastpage}
\end{document}